\documentclass[sigplan,nonacm]{acmart}

\setcopyright{none}
\makeatletter
\@ifundefined{hyxmp@parse@acmart}{}{%
  \apptocmd{\hyxmp@parse@acmart}{%
    \hypersetup{pdfpublisher={},pdfpubtype={},pdfvolumenum={},pdfissuenum={}}%
  }{}{\PackageError{weaver}{Could not clear ACM publication metadata}{}}
}
\makeatother
\usepackage{subfig}
\usepackage{pifont}
\usepackage{xcolor}
\usepackage{booktabs}
\usepackage{multirow}
\usepackage{threeparttable}
\usepackage{algorithm}
\usepackage{algpseudocode}

\newcommand{\sysname}{Weaver}

\newcommand{\etitle}[1]{\vspace{1ex}\noindent{\em\underline{#1}}}
\definecolor{navyblue}{RGB}{0, 0, 128}

\AtBeginDocument{}

\title{\sysname{}: Unlocking Blockchain Transaction Parallelism through State-Decoupled Symbolic Execution}
\author{Ru Cheng}
\affiliation{%
  \institution{City University of Hong Kong}
  \city{}
  \country{}
}
\email{ru.cheng@my.cityu.edu.hk}

\author{Maofan Yin}
\affiliation{%
  \institution{Lyquor Labs}
  \city{}
  \country{}
}
\affiliation{%
  \institution{UC Santa Barbara}
  \city{}
  \country{}
}
\email{ted@lyquor.xyz}

\author{Liangrong Chen}
\affiliation{%
  \institution{City University of Hong Kong}
  \city{}
  \country{}
}
\email{liangrong.chen@my.cityu.edu.hk}

\author{Lizan Zhou}
\affiliation{%
  \institution{Lyquor Labs}
  \city{}
  \country{}
}
\email{lizan@lyquor.xyz}

\author{Chenyuan Wu}
\affiliation{%
  \institution{City University of Hong Kong}
  \city{}
  \country{}
}
\email{chenyuan.wu@cityu.edu.hk}

\renewcommand{\shortauthors}{Cheng et al.}

\begin{document}

\begin{abstract}
Transaction execution remains a major performance bottleneck in blockchains. Existing parallel execution engines are still blocked by slow state accesses on each transaction's critical path, creating a hidden loss of parallelism. This paper presents \sysname{}, a state-decoupled transaction processing paradigm that enables execution flow to proceed in the absence of concrete state values. \sysname{} represents states as lightweight symbolic references and records subsequent computation as symbolic expressions while concrete values are fetched on the fly. To handle state-dependent control flow, \sysname{} predicts the most likely branch direction through hierarchical symbolic pattern matching. \sysname{} further introduces expression-granular scheduling that evaluates each symbolic expression as soon as its state dependencies are resolved, rather than waiting for the transaction's entire read set, fully exploiting fine-grained intra-transaction parallelism. Evaluation on real Ethereum blocks and synthetic workloads spanning four representative dApps shows that \sysname{} delivers 41\%--63\% higher throughput than state-of-the-art parallel execution engines and up to 7.5$\times$ the throughput of serial execution.
\end{abstract}


\maketitle
\pagestyle{plain}

\section{Introduction}\label{sec:introduction}
Smart contracts have expanded blockchain applications beyond simple cryptocurrency transfers, enabling a diverse range of decentralized applications (dApps), ranging from decentralized finance (DeFi)~\cite{aft2022defi,sp2020flashboys} and on-chain games~\cite{usenixsecurity2023metaversegames,usenixsecurity2018erays} to decentralized physical infrastructure networks (DePINs)~\cite{atc2023internetcomputer,usenixsecurity2022scraps,icde2023saguaro}. To finalize each block, validators must execute its ordered transactions and derive an identical post-state. With recent advances in consensus and transaction dissemination~\cite{eurosys2022narwhal,ccs2022bullshark,ndss2025mysticeti,nsdi2025shoalpp}, transaction execution has emerged as a major scalability bottleneck, as modern smart-contract workloads involve increasingly complex computation together with frequent accesses to persistent state.

Existing systems have made substantial progress in accelerating blockchain execution by exploiting inter-transaction parallelism. Dependency-graph-based approaches~\cite{icpp2023blockpilot, icdcs2019parblockchain} identify transaction conflicts in advance and schedule non-conflicting transactions concurrently. Optimistic execution engines~\cite{icse2022occda,sigmod2019fabric,sigmod2020transactional,tkde2021concurrency,vldb2021slimchain} instead execute all transactions concurrently and validate their effects against the prescribed order afterwards. Software transactional memory (STM)-based systems~\cite{ppopp2023blockstm,icdcs2023dmvcc} dynamically discover dependencies during execution and re-execute conflicting transactions. More recent operation-level techniques~\cite{vldb2024spectrum,eurosys2025parallelevm} further reduce re-execution overhead by recovering only the affected operations rather than entire transactions.

Despite these advances, existing execution engines still share a fundamental limitation: while they can overlap computations across different transactions, state accesses (i.e., operations that read or update persistent blockchain state) remain a non-negligible blocking source within each transaction. Specifically, in an Ethereum Virtual Machine (EVM) transaction, execution is organized as a stream of bytecode instructions, where instructions that trigger state accesses are interleaved with those performing computations over different states as well as state-independent computations. These engines require concrete operands to advance a transaction's instruction stream, lacking a mechanism to carry unresolved computations forward while continuing subsequent execution. A pending state read therefore stalls the entire instruction stream, even when subsequent parts of the transaction do not immediately require its value or derived results, leaving intra-transaction parallelism unexploited.

To this end, we propose \sysname{}, a state-decoupled transaction processing paradigm that enables execution flow to proceed in the absence of concrete state values. Our key insight is that many instructions only need to propagate or transform a state-derived value, rather than evaluate its concrete result immediately. Following this observation, \sysname{} adopts \textit{lazy state concretization}: when a requested state is unavailable, it issues the backend access asynchronously and represents the pending value with a lightweight symbolic reference. Subsequent computations over the reference are recorded as state-derived symbolic expressions until concrete evaluation becomes necessary. In this way, \sysname{} separates instruction execution progress from immediate state retrieval, moving state-access latency off the critical path whenever the corresponding value is not yet required.

State-decoupled execution introduces a key challenge in handling control flow. While many instructions can operate on symbolic values, conditional branches require a definite outcome of the symbolic condition before execution can proceed. Resolving such conditions using concrete values would place state access back on the critical path, whereas incorrect decisions may divert execution from the actual path and trigger excessive recovery. \sysname{} employs hierarchical branch prediction to preserve the decoupling while balancing this trade-off. Instead of relying on a fixed prediction context, \sysname{} begins with coarse and reusable symbolic patterns, and progressively incorporates more execution context only when earlier levels remain inconclusive. This design enables stable branches to be predicted quickly at upper levels with low overhead and broad reuse, while deeper levels capture more context-sensitive and evolving branch behaviors, thereby improving prediction accuracy and deferring forced concretization for more branches.

Symbolic representation also brings new opportunities. Computational dependencies that are normally hidden within a transaction's sequential instruction stream are exposed, enabling \sysname{} to move beyond transaction-granular scheduling. Specifically, rather than waiting for all inputs of a transaction to become available, \sysname{} decomposes its state-derived computation into independently evaluable expressions according to their runtime-derived dependencies. Each expression can be scheduled once its required states become available and reflect the preceding updates prescribed by the block order, without waiting for unrelated states in the same transaction. This expression-granular scheduling naturally exploits intra-transaction parallelism, without requiring explicit declarations of independent computation.

We implement \sysname{} on Geth and compare it against native serial execution and state-of-the-art execution engines, including OCC~\cite{icse2022occda}, BlockSTM~\cite{ppopp2023blockstm}, and ParallelEVM~\cite{eurosys2025parallelevm}. We replay 50k consecutive Ethereum mainnet blocks from September 2025 and use synthetic workloads spanning four representative dApps. The results show that \sysname{} achieves 41--63\% higher throughput than state-of-the-art systems and up to 7.5$\times$ the throughput of serial execution.

Overall, this paper makes the following contributions:
\begin{itemize}
    \item We identify a loss of parallelism between state access and instruction execution, and present \sysname{}, a state-decoupled execution paradigm based on lightweight symbolic execution.

    \item We design lightweight symbolic representations and hierarchical branch prediction to enable execution over unresolved state values with low overhead.

    \item We design expression-granular scheduling that leverages runtime-derived computational dependencies to exploit fine-grained intra-transaction parallelism.

    \item We implement a prototype of \sysname{} and evaluate it using real Ethereum blocks and synthetic workloads, demonstrating its throughput improvements over state-of-the-art execution engines.
\end{itemize}

\section{Background \& Motivation}\label{sec:background&motivation}
\subsection{Smart Contract Execution}
Smart contracts are programs deployed on blockchains to implement application logic that may read or update persistent blockchain state. Ethereum executes smart contracts through the Ethereum Virtual Machine (EVM), a stack-based virtual machine operating on bytecode instructions. During execution, the EVM maintains a program counter and an operand stack, while persistent contract data resides in storage across transactions. Instructions such as \texttt{SLOAD} and \texttt{SSTORE} directly access persistent contract state. \texttt{SLOAD} reads a value from storage and places it onto the execution stack, while \texttt{SSTORE} writes a stack-derived value back to storage.

Once loaded onto the stack, a state value may participate in a sequence of subsequent instructions before eventually contributing to another state update. Literal constants are similarly introduced onto the stack through instructions such as \texttt{PUSH} and participate in subsequent computations. Arithmetic instructions consume stack operands and push their results back, allowing values derived from storage to be progressively transformed or combined with other operands, while stack management instructions such as \texttt{DUP} and \texttt{SWAP} replicate or reposition them for later use. Stack values may also determine control flow, since \texttt{JUMPI} consumes a stack operand as its condition and selects the subsequent execution path accordingly. Beyond stack operations, intermediate values can be written to temporary memory and later recovered for further computation, extending their lifetime beyond immediate instruction dependencies. Contract calls further propagate values across call frames, allowing computation results originating from one contract to span multiple contract invocations.

\subsection{Blockchain State Access}

Ethereum maintains an authenticated world state recording persistent account state and contract storage. At the protocol level, account states are organized in a Merkle Patricia Trie (MPT), where each account record contains its balance and nonce alongside a code hash and storage root. For a contract account, the storage root identifies another MPT holding its persistent storage slots. A contract-state lookup conceptually traverses the global state trie to locate the account before resolving the slot from its storage trie. These authenticated structures support committed-state verification while requiring multi-level traversal during state retrieval.

At a lower layer, blockchain clients persist trie nodes and state objects in LSM-tree-based key-value stores. Such stores favor write efficiency, while reads may traverse multiple storage levels before reaching persistent data. Clients therefore employ in-memory caches to serve frequently accessed states and reduce backend lookups. As blockchain state grows and access locality shifts, however, caching cannot eliminate backend retrieval altogether. During conventional EVM execution, a state-access instruction such as \texttt{SLOAD} must return a concrete value to the operand stack before subsequent computation can proceed. Consequently, state-retrieval latency can directly translate into transaction-execution stalls.

\subsection{Motivation}

\begin{figure}[t]
    \centering
    \subfloat[500 blocks]{
        \includegraphics[width=0.22\textwidth]{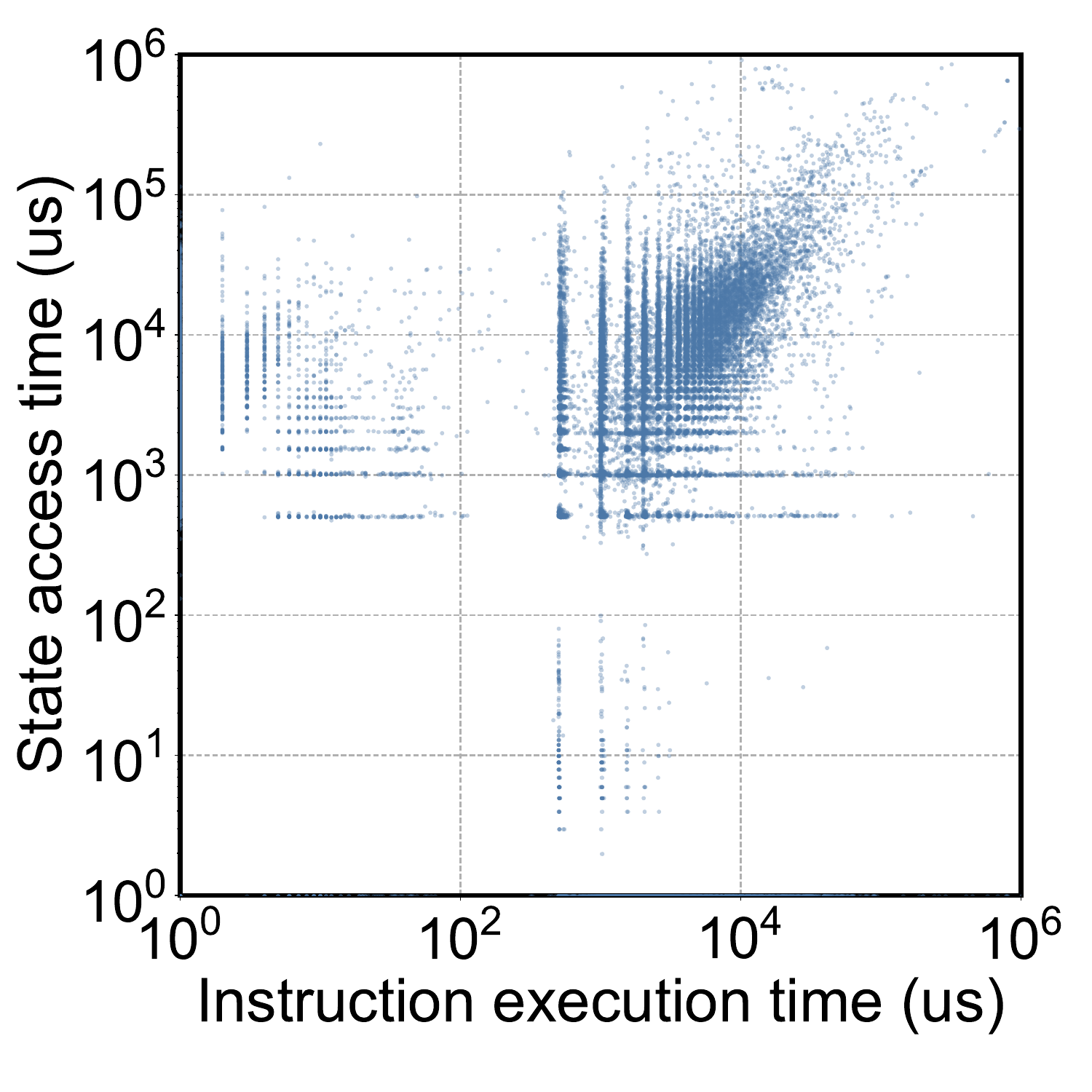}
        \label{fig:state-access-distribution-small}
    }
    \subfloat[10,000 blocks]{
        \includegraphics[width=0.22\textwidth]{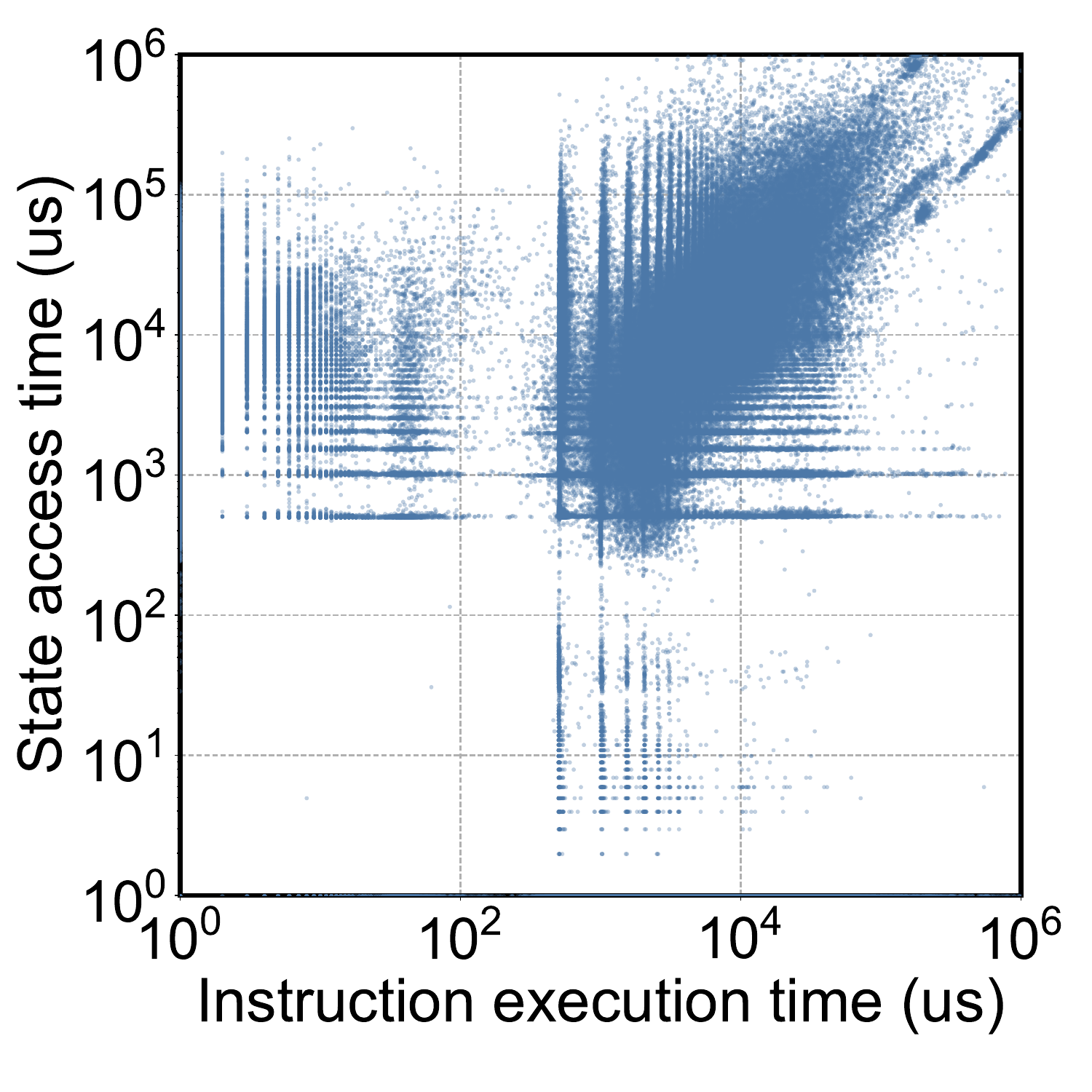}
        \label{fig:state-access-distribution-large}
    }
    \caption{Per-transaction time spent on state access and other instruction execution under two replay settings. Each point represents one transaction.}
    \label{fig:state-access-distribution}
    \Description{Two scatter plots for 500 and 10,000 blocks. Each point represents a transaction, with instruction-execution time on the horizontal axis and state-access time on the vertical axis. Both axes use logarithmic scales.}
\end{figure}

To characterize the impact of state access, we replay a series of Ethereum blocks and measure the time each transaction spends on state access and executing all other instructions. We use the original Geth implementation and evaluate two database configurations populated with the states required to replay 500 and 10,000 blocks, respectively. The remaining testbed configurations are described in Section~\ref{subsec:experimentalsetup}. Figure~\ref{fig:state-access-distribution} plots each transaction as one point and reveals two key observations. First, state access accounts for a non-negligible portion of transaction execution time, leaving substantial opportunity to overlap state retrieval with bytecode computation rather than placing every read directly on the critical path. Second, state-access time varies considerably across transactions, even among those with similar instruction-execution times. This variability suggests that the states required by different transactions, or by different computational parts within the same transaction, may become available at substantially different times. Together, these observations motivate decoupling instruction execution from state access and scheduling subsequent expression evaluations more flexibly according to the readiness of their respective state inputs.

\subsection{Challenges}
Building a practical state-decoupled execution engine, however, requires overcoming three main challenges.

\noindent\textbf{Efficient representation and propagation of symbolic values.} Symbolic execution must remain close to native EVM execution performance while recording computations over unresolved state values. A naive implementation can incur substantial allocation and bookkeeping overhead as symbolic expressions are constructed and propagated through the EVM stack, memory, and contract call frames. Since these operations are performed on the real-time transaction-processing path, their overhead may offset the benefits of state decoupling. Minimizing these costs while faithfully preserving symbolic value representations and their computational dependencies remains a key challenge.

\noindent\textbf{Accurate control-flow decisions over symbolic expressions.}
Although symbolic execution can advance through most bytecode operations without concrete state values, conditional branches require a definite control-flow decision before execution can proceed. If the condition of a \texttt{JUMPI} instruction depends on a symbolic expression, resolving it with concrete values would place state access back on the critical path, whereas an incorrect decision may divert execution from the actual path and trigger recovery. The system therefore requires an efficient mechanism that accurately predicts branch directions from symbolic conditions and execution context without relying on concrete state values.

\noindent\textbf{Incompatibility with existing concurrency control mechanisms.} Existing concurrency control mechanisms mainly coordinate execution at transaction granularity, whereas state decoupling exposes finer-grained computations within each transaction. Different such computations may become executable at different times, depending on the readiness of their own state inputs. The system must exploit such partial readiness across and within transactions while preserving the state dependencies imposed by the block order.

\section{\sysname{} Design}\label{sec:design}
We present \sysname{}, a state-decoupled high-performance blockchain execution engine. At a high level, \sysname{} adopts an execution-and-evaluation design: each transaction is executed symbolically alongside asynchronous state access to produce fine-grained state-derived expressions that are concurrently evaluated as their required states become available.

\subsection{Overview}\label{subsec:overview}

\begin{figure}[t]
  \centering
  \includegraphics[width=0.98\linewidth]{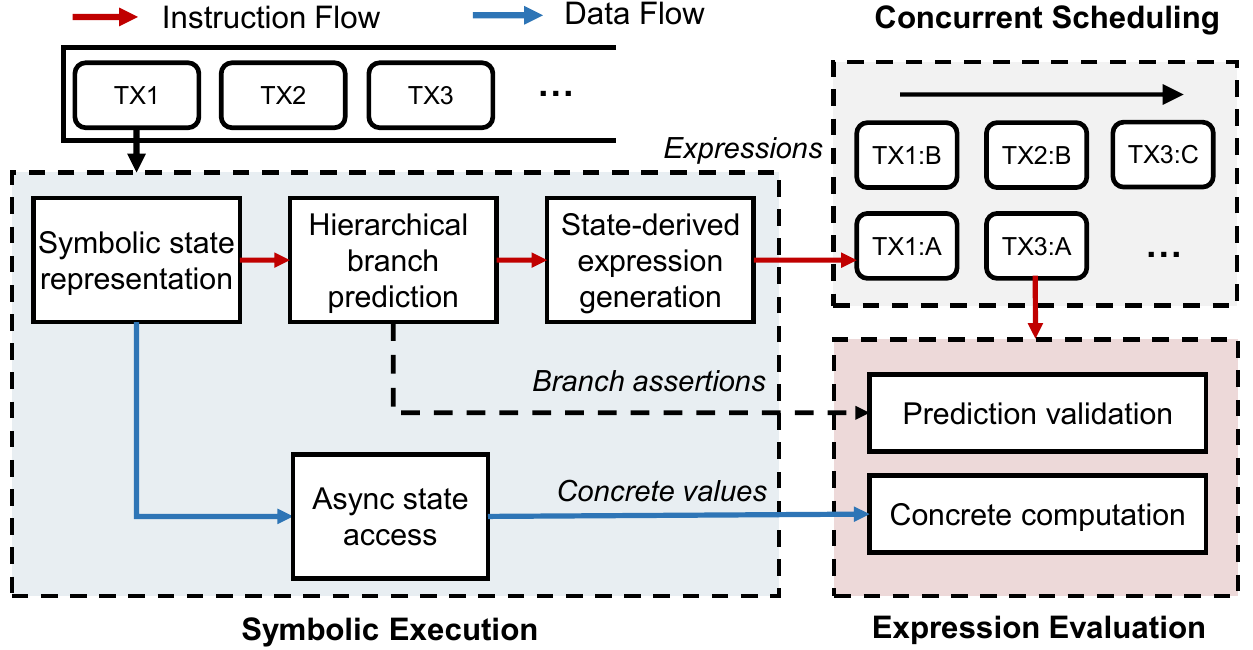}
  \caption{\sysname{} overview. Symbolic execution proceeds alongside asynchronous state access to produce state-derived expressions and branch assertions. Concurrent scheduling dispatches expressions for concrete evaluation as their required states become available and up-to-date.}\label{fig:overview}
  \Description{}
\end{figure}

Fig.~\ref{fig:overview} provides an overview of \sysname{}. When transaction execution encounters an unavailable blockchain state, \sysname{} issues an asynchronous request to retrieve it from the state backend. In parallel, it represents the pending value with a lightweight symbolic reference and continues execution symbolically. As execution proceeds, \sysname{} propagates these references through the stack, memory, and call frames, incrementally constructing state-derived expressions that capture how written states depend on previously read states. State-dependent branches require special handling because they determine the subsequent execution path. For example, a \texttt{JUMPI} instruction may use a symbolic value as its condition. \sysname{} handles such branches through hierarchical branch prediction, which derives branch decisions from symbolic conditions and execution context. Predicted outcomes are recorded as branch assertions for later validation. The details of symbolic value representation and branch prediction are presented in Sec.~\ref{subsec:execution} and Sec.~\ref{subsec:prediction}, respectively.

After symbolic execution of a transaction, \sysname{} produces its read/write sets together with expressions capturing the computational dependencies from relevant read states to each state-dependent write. Since state accesses and preceding updates for different states complete asynchronously, these expressions may become evaluable at substantially different times. \sysname{} therefore treats each expression as an independently schedulable task that can proceed once its dependent states become ready, without waiting for the transaction's entire read set. During evaluation, \sysname{} substitutes concrete values of these states into the recorded expression to derive the corresponding write. Predicted branch outcomes are retained as assertions and validated under the same scheduling framework by evaluating their condition expressions, with failed validations triggering recovery of affected computations. The details of expression-granular scheduling are given in Sec.~\ref{subsec:schedule}. \sysname{} further improves efficiency by suppressing redundant state accesses and reusing valid intermediate results, while avoiding full-transaction fallback from branch mispredictions, as detailed in Sec.~\ref{subsec:optimizations}.

\subsection{Symbolic Value Representation}\label{subsec:execution}
\sysname{} represents state-dependent values using state references and state-derived expressions. A state reference identifies a specific blockchain state item, while a state-derived expression captures computations over one or more such references. When a transaction first loads a state item onto the stack, instead of waiting for the concrete value, \sysname{} introduces the corresponding reference as a symbolic operand, while initiating an asynchronous request in the backend for its value. Subsequent computational instructions construct expressions over this reference without waiting for the state request to complete. Values independent of dynamic state retain their native representations, and operations on them are processed directly by the execution engine.

\noindent\textbf{Expression construction.}
\sysname{} incrementally constructs symbolic expressions as instructions consume and produce values, covering three cases.
\ding{182} \textbf{Expression extension.} For binary operations, when one operand is a symbolic value while the other is concrete, \sysname{} extends the former to form a new expression. For instance, applying \texttt{ADD} to state reference $s_X$ and constant $5$ produces $e_1=s_X+5$. A subsequent multiplication may further extend it to $e_2=2e_1$.
\ding{183} \textbf{Expression composition.} When both operands are symbolic, \sysname{} combines them using the corresponding operator. For example, applying \texttt{SUB} to $e_2$ and an expression $e_3$ derived from $s_Y$ produces $e_4=e_2-e_3$. The result retains dependencies on both inputs, allowing multi-state computations to remain symbolic.
\ding{184} \textbf{Expression sharing.} An intermediate expression may also participate in multiple computations. Specifically, \texttt{DUP} duplicates a symbolic value across stack elements, allowing each copy to be independently consumed by subsequent instructions. Duplicating $e_1$, for example, allows one copy to undergo further computation while another is used for an integrity check. \sysname{} retains a shared representation of the common intermediate expression and separately records the additional operations for each derived computation, leaving the original expression unchanged.

\noindent\textbf{Expression propagation.} Beyond stack-local operations, \sysname{} also propagates symbolic values across other EVM data containers and contract-call boundaries. Specifically, when \texttt{MSTORE} writes a symbolic value to memory, \sysname{} associates its symbolic representation with the corresponding byte range so that a subsequent \texttt{MLOAD} from that range can restore the symbolic value onto the stack. Across contract calls, \sysname{} propagates symbolic representations alongside call data and return data, allowing the receiving call frame to recover the corresponding symbolic values.

\noindent\textbf{Compact expression representation.} To keep expression construction and propagation lightweight, in practice, \sysname{} keeps successive expression extensions within a single compact entry whenever possible, introducing new entries only when expressions are composed or forked for distinct continuations. Each symbolic stack value therefore carries only a lightweight \textit{entry.offset} identifier to locate its symbolic representation, avoiding repeated storage or copying of complete expressions during stack operations such as \texttt{DUP} and \texttt{SWAP}. For memory and call/return data, \sysname{} similarly maintains lightweight byte-range metadata to preserve symbolic values during cross-container propagation.

\noindent\textbf{Symbolic execution outputs.} Upon completion, symbolic execution produces read and write sets for each transaction. However, unlike the conventional write set that only records which states are updated, \sysname{} associates each state-dependent write with its corresponding expression. These expressions preserve their computational dependencies on one or more state values, identifying the inputs required for subsequent evaluation. \sysname{} then schedules each write expression for concrete evaluation based on the readiness of its required state values (detailed in Sec.~\ref{subsec:schedule}).

\begin{figure}[t]
  \centering
  \includegraphics[width=1.00\linewidth]{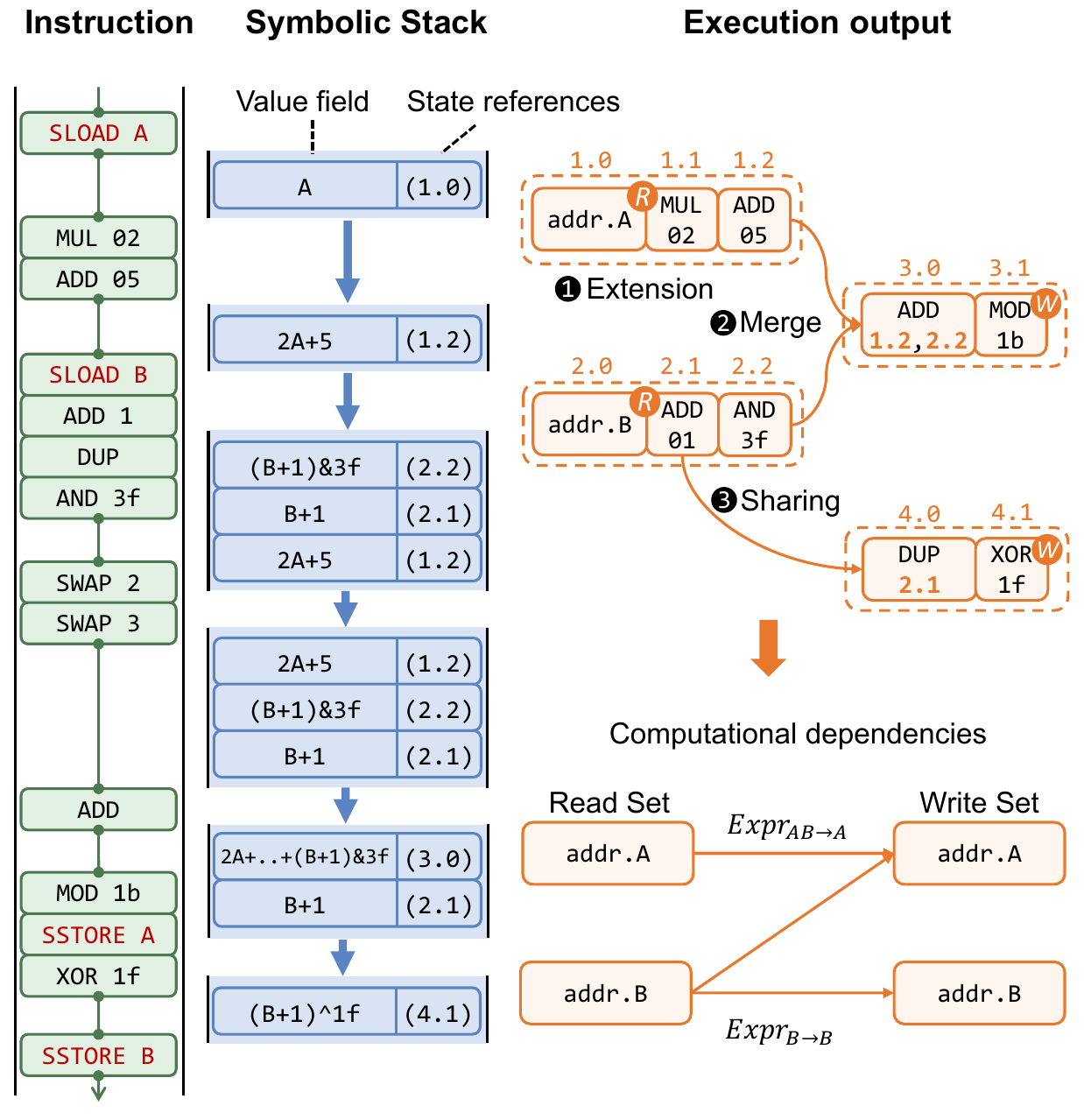}
  \caption{Symbolic value representation and expression construction. The left, middle, and right sides show the executed EVM instructions, symbolic stack, and expression construction with resulting read/write dependencies, respectively. The tuple $\mathit{entry.offset}$ identifies an expression entry and its operation offset, allowing symbolic operands to carry only a compact reference rather than the complete expression.}
  \label{fig:symbolic_execution}
  \Description{}
\end{figure}

\noindent\textbf{Example.}
Fig.~\ref{fig:symbolic_execution} illustrates symbolic execution for a transaction that reads and updates state items $A$ and $B$. The \texttt{SLOAD} instructions introduce state references $s_A$ and $s_B$ onto the symbolic stack. For $A$, successive \texttt{MUL} and \texttt{ADD} operations construct the expression $2s_A+5$, referenced by 1.2 on the stack. For $B$, \sysname{} first constructs $s_B+1$, which \texttt{DUP} allows two subsequent computations to share. One copy is extended by \texttt{AND}, while the other retains the original expression. After \texttt{SWAP} rearranges the operands, \texttt{ADD} combines the expressions derived from $A$ and $B$, and \texttt{MOD} further extends the result before the updated $A$ is written back. The retained copy of $s_B+1$ is subsequently extended by \texttt{XOR} and written to $B$. Finally, the execution produces two write expressions in the write set, one for $A$ and one for $B$: the expression for $A$ depends on the original values of both $A$ and $B$, whereas the expression for $B$ depends only on the original value of $B$.

\subsection{Hierarchical Branch Prediction}\label{subsec:prediction}
\noindent\textbf{State-dependent branches.}
Symbolic execution allows computational instructions to proceed by propagating state references and constructing state-derived expressions. Conditional control-flow instructions, however, require a clear branch direction to continue processing along the correct path. For example, the \texttt{JUMPI} instruction decides whether to enter a branch by evaluating the value on the stack representing the condition, which may not be available at that point and is still presented as a state-derived expression. Resolving every conditional value when encountering a branch would reintroduce state access as a blocking source, while choosing a branch without sufficient evidence may cause symbolic execution to deviate from the correct execution path, resulting in excessive overhead due to re-execution. A predictor is therefore needed to accurately select branch directions consistent with the paths that transactions ultimately follow during commit. To this end, \sysname{} devises a hierarchical prediction mechanism by exploiting recurring patterns in previously constructed expressions.

\noindent\textbf{Hierarchical pattern matching.}
\sysname{} predicts branch directions by matching symbolic patterns against previously validated outcomes. Matching the full execution context, however, requires substantial history storage and yields a low hit rate. Our first insight is that the shapes of symbolic expressions at branching points often capture recurring semantics across contracts and program locations. For example, balance-sufficiency branches commonly follow the form \texttt{ISZERO(LT(var,const))}, while arithmetic-overflow checks follow \texttt{ISZERO(LT(ADD(var,const),var))}. Other conditional branch semantics likewise exhibit their respective recurring shapes. \sysname{} therefore uses abstracted expression shapes as the initial matching granularity and predicts the dominant direction based on historical outcomes, enabling fast and efficient branch prediction.

However, certain branches with the same expression shape may still favor different directions depending on finer-grained execution context, such as contract identity or the branch's temporal behavior within a transaction or block. In such cases, matching only on expression shape may compromise prediction accuracy. This motivates our hierarchical design: \sysname{} incorporates additional context whenever a coarser pattern lacks sufficient confidence. Specifically, if the current level fails to produce a prediction above a confidence threshold (e.g., 95\%), \sysname{} proceeds to the next level, augmenting the matching key with additional context. This hierarchical refinement continues until the matched history supports a sufficiently confident prediction, thereby retaining efficient coarse-grained matching for confident cases while refining ambiguous ones with additional context.

\noindent\textbf{Prediction rules.}
Prediction rules also become progressively refined as the hierarchy incorporates additional context. At coarser levels, \sysname{} simply uses the historically dominant direction as the default prediction rule. At deeper levels, the additional context enables finer-grained rules. For example, some branches may follow recurring cyclic patterns across successive occurrences, whereas others may take one direction for one or several occurrences and then flip to another stable direction within a transaction or block. Accordingly, when branch sequence information is incorporated, \sysname{} identifies such temporal patterns from validated history and applies the corresponding prediction rules.

\begin{figure}[t]
  \centering
  \includegraphics[width=0.98\linewidth]{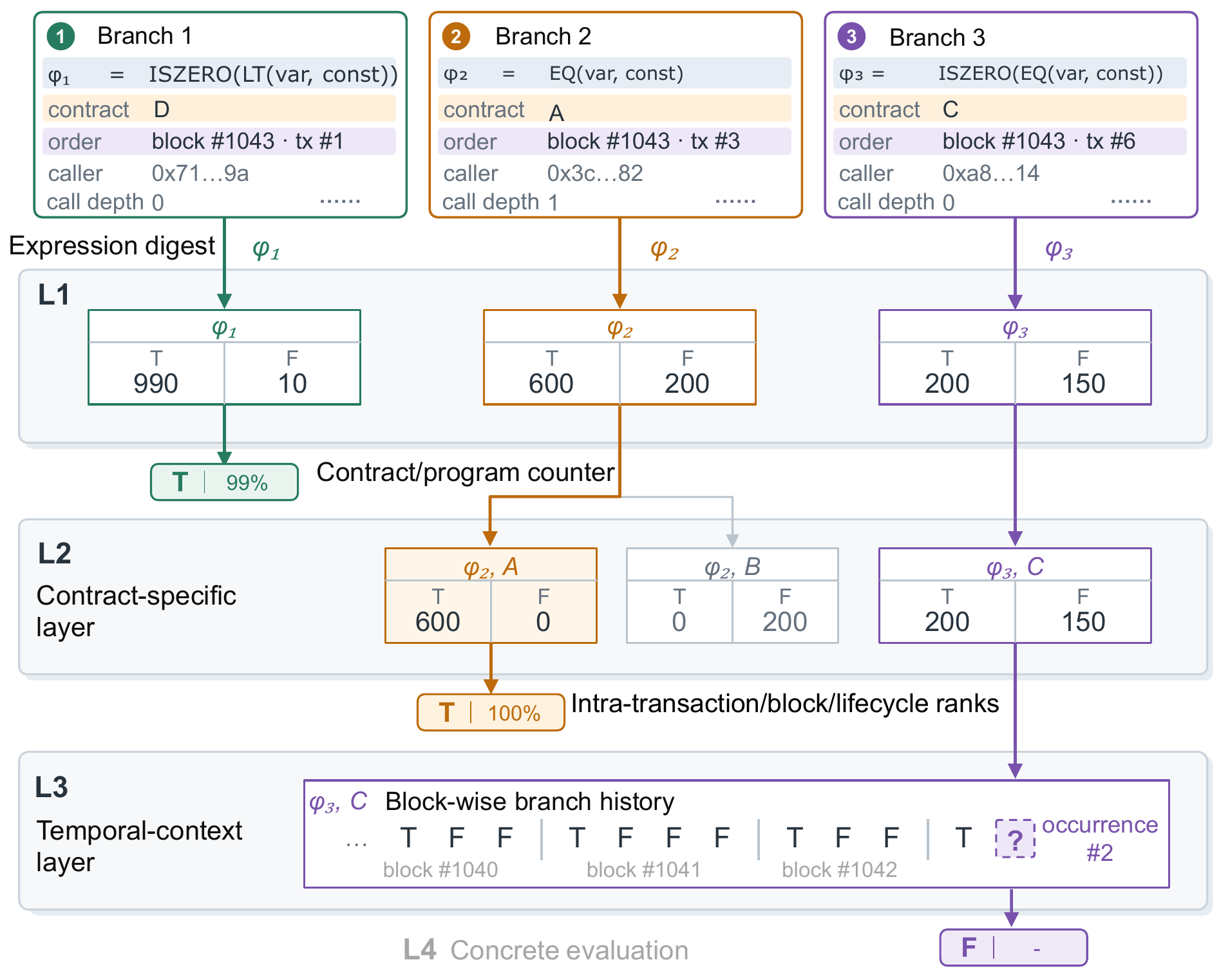}
  \caption{A simplified example of hierarchical branch prediction. Branches 1--3 are resolved at L1--L3, respectively, as the context is progressively refined from expression shape to contract-specific and temporal information.}
  \label{fig:branch_prediction}
\end{figure}

Specifically, \sysname{} employs the following four-level hierarchy for branch matching and prediction:

\etitle{L1 (Expression-shape prediction)}.
This level matches symbolic expression shapes across contracts and program locations, capturing recurring patterns whose outcomes tend to favor a stable direction. By sharing validated history across matching expressions, it provides compact and reusable predictions. \sysname{} returns the dominant direction when its confidence meets the predefined threshold. For example, Branch 1 in Fig.~\ref{fig:branch_prediction} matches a shape whose history strongly favors \texttt{true}, and is therefore resolved directly at L1.

\etitle{L2 (Contract-specific prediction)}.
When the aggregate history of an expression shape has no clear dominant direction, individual contracts may still exhibit strong local biases. \sysname{} therefore augments the matching key with the contract address and branch location, separating histories for branches that share the same shape but follow different contract-specific behaviors and deriving predictions from this finer context. In Fig.~\ref{fig:branch_prediction}, Branch 2 is ambiguous at L1, while its history in the matched contract reflects a strong bias toward \texttt{true}, thus delivering a confident prediction at L2.

\etitle{L3 (Temporal-context prediction)}.
This level further incorporates temporal information of a branch into the prediction context, namely, its occurrence order within an execution scope such as the current transaction, block, or whole contract history. This enables \sysname{} to capture how branch behavior evolves across successive occurrences. Some branches change direction at particular occurrences, while others may follow recurring cyclic patterns. Such temporal-context prediction enables confident predictions when the aggregate history lacks a clear dominant direction. In Fig.~\ref{fig:branch_prediction}, Branch 3 remains ambiguous at L1 and L2, while its temporal context at L3 reveals a consistent flip after the first occurrence within each block, thus yielding a confident \texttt{false} prediction.

\etitle{L4 (Concrete evaluation)}.
If no earlier level provides sufficient confidence, \sysname{} falls back to concrete evaluation once the required state values become available. Rather than blocking the execution thread while waiting, \sysname{} can switch to another runnable transaction and resume the suspended one later. We detail this task-switching mechanism in Sec.~\ref{subsec:schedule}.

\noindent\textbf{Prediction validation and feedback.}
\sysname{} records each predicted direction as a branch assertion and validates it once the condition expression can be concretely evaluated with the latest state values. The resulting validated outcomes are fed back to update direction statistics at L1/L2 and temporal-pattern statistics at L3 to guide subsequent predictions. When a coarse entry remains inconclusive, \sysname{} creates finer entries for the observed context, which become eligible for prediction only after accumulating sufficient observations to meet the corresponding confidence threshold. To keep the predictor adaptive and compact, \sysname{} gradually decays histories that remain inconclusive across all levels over a sufficiently large observation window, or become inactive before any level establishes a confident pattern.

\subsection{Expression-Granular Scheduling}\label{subsec:schedule}
Symbolic execution decomposes a transaction into individual expressions, each representing a distinct computation within the transaction. Since each expression records its own state dependencies, expressions can be scheduled for evaluation independently as their required states become available, without waiting for unrelated reads. Therefore, \sysname{} designs a fine-grained concurrency control mechanism that schedules transaction evaluation at expression granularity according to state readiness.

\noindent\textbf{Task model.}
In \sysname{}, each transaction comprises a symbolic execution task and several expression-evaluation tasks. For transaction $\mathit{tx}_i$ at position $i$ in the block, task $E_i$ executes its instructions symbolically and produces write expressions and branch assertions. Let $\mathcal W_i$ denote the set of states written by $\mathit{tx}_i$. For each $w\in\mathcal W_i$, task $V_i^w$ evaluates the corresponding write expression $e_i^w$ using the values of the states in its dependency set $\mathcal R_i^w$ and applies the resulting update to $w$. For consistency and simplicity of our model, branch-prediction validations are also treated as evaluation tasks over state-dependent condition expressions. Each such validation, however, further compares its result with the recorded branch assertion, and a mismatch may invalidate relevant tasks and trigger re-execution.

\noindent\textbf{State readiness.}
For an evaluation task $V_i^X$ to be considered ready, each of its required states $Y\in\mathcal R_i^X$ must satisfy the following three conditions:

\etitle{C1 (Value availability)}.
The corresponding asynchronous access to the state backend must have finished, rendering the concrete value of $Y$ available for expression evaluation. However, the retrieved value may not yet reflect updates from preceding transactions. \sysname{} therefore further enforces conditions \textit{C2} and \textit{C3} to ensure that all relevant preceding updates have been identified and applied, respectively.

\etitle{C2 (Dependency discovery)}.
All symbolic-execution tasks of preceding transactions within the block, i.e., $\{E_j\mid j<i\}$, must have been completed to allow \sysname{} to identify all preceding expressions that write to $Y$. Specifically, unfinished tasks may still generate write expressions for $Y$, introducing a dependency that has not yet been identified. Completing all such tasks establishes the full set of preceding updates to $Y$ for the subsequent freshness check.

\etitle{C3 (Value freshness)}.
All expression-evaluation tasks for preceding transactions within the block that update $Y$, i.e., $\{V_j^Y\mid j<i\}$, must have been completed to ensure the freshness of $Y$ when evaluating $V_i^X$. While \textit{C2} identifies all preceding updates to $Y$, \textit{C3} ensures that these updates have been evaluated and applied in block order. Their completion guarantees that all preceding updates identified in \textit{C2} have been incorporated into the value used by $V_i^X$, thereby preserving serial equivalence to the predetermined transaction order.

Once considered ready, the expression can be scheduled for evaluation, even if other evaluation tasks from preceding transactions remain unfinished or other expressions from the same transaction are not yet ready.

\algblockdefx[Upon]{Upon}{EndUpon}
  [1]{\textbf{upon} #1 \textbf{do}}
  {\textbf{end upon}}

\begin{algorithm}[t]
\caption{Expression-Granular Task Scheduling}
\label{alg:runtime-scheduling}
\small
\renewcommand{\baselinestretch}{0.92}\selectfont

\begin{algorithmic}[1]

\State \textbf{Input:} ordered transactions
$\mathcal{T}=\langle tx_1,\ldots,tx_n\rangle$

\State \textbf{Data:}
$L_{\mathrm{ready}}$: ordered list of tasks ready for execution;
\Statex
$L_{\mathrm{wait}}[s]$: tasks waiting on state $s$;
\Statex
$L_{\mathrm{write}}[s]$: transactions that write state $s$.

\Statex {\color{navyblue}\textbf{$\triangleright$ Initialization}}

\State $L_{\mathrm{ready}} \gets \langle E_1,\ldots,E_n\rangle$
\State $L_{\mathrm{wait}}[s], L_{\mathrm{write}}[s]
       \gets \varnothing,\ \forall s$

\Statex {\color{navyblue}\textbf{$\triangleright$ Task Execution}}

\For{each worker \textbf{in parallel}}
    \While{$task \gets L_{\mathrm{ready}}.pop()$}
        \If{$task = E_i$}
            \State $(\mathcal{R}_i,\mathcal{W}_i) \gets E_i.execute()$
            \For{each $w\in\mathcal{W}_i$}
                \State Register $i$ in $L_{\mathrm{write}}[w]$
                \State Register $V_i^w$ in $L_{\mathrm{wait}}[s]$, $\forall s\in\mathcal{R}_i^w$
                \State \Call{CheckReady}{$V_i^w$}
            \EndFor
        \Else
            \State Evaluate $V_i^w$ and apply its update to $w$
            \State Recheck affected tasks in $L_{\mathrm{wait}}[w]$
        \EndIf
    \EndWhile

\EndFor

\Statex {\color{navyblue}\textbf{$\triangleright$ Readiness Check}}
\Procedure{CheckReady}{$V_i^w$}
    \For{each $s\in\mathcal{R}_i^w$}
        \State Find the latest preceding writer of $s$
               in $L_{\mathrm{write}}[s]$
        \If{the preceding write to $s$ has completed}
            \State Remove $V_i^w$ from $L_{\mathrm{wait}}[s]$
        \EndIf
    \EndFor
    \If{$V_i^w$ has no unresolved dependencies}
        \State Add $V_i^w$ to $L_{\mathrm{ready}}$
    \Else
        \State Register $V_i^w$ in $L_{\mathrm{wait}}[s]$, $\forall$ unresolved $s$
    \EndIf
\EndProcedure

\Statex {\color{navyblue}\textbf{$\triangleright$ State-Access Completion}}

\Upon{\textsc{StateAccessComplete}($s$)}
    \State Mark the initial access to $s$ as completed
    \State \Call{CheckReady}{$L_{\mathrm{wait}}[s].first()$} \textbf{if} $L_{\mathrm{wait}}[s]\neq\varnothing$
\EndUpon

\end{algorithmic}
\vspace{0.5em}
\end{algorithm}

\begin{figure*}[t]
  \centering
  \includegraphics[width=\linewidth]{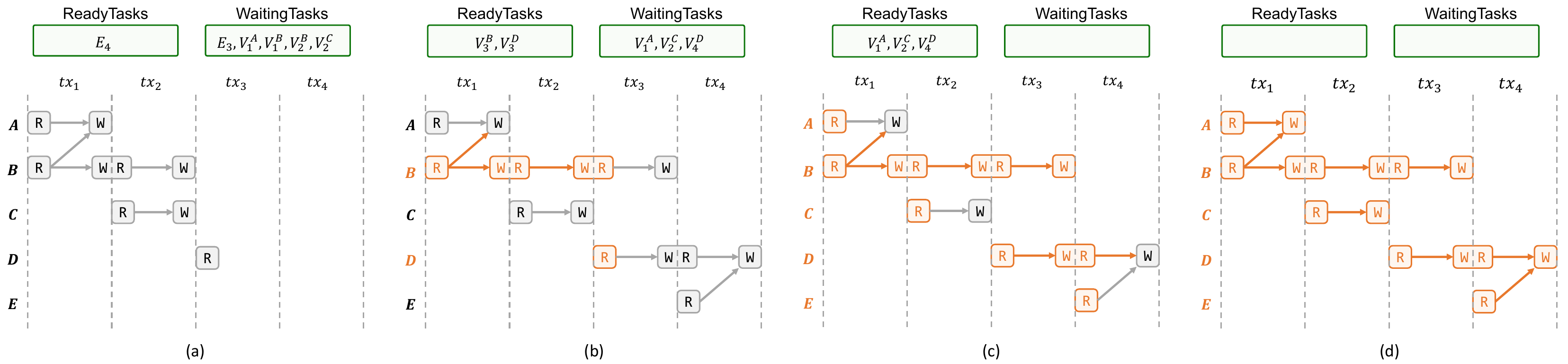}
  \caption{Example of expression-granular scheduling under state readiness. $A$--$E$ denote blockchain states, $E_i$ denotes the symbolic-execution task of transaction $\mathrm{tx}_i$, and $V_i^X$ denotes the expression-evaluation task that computes the update to state $X$ for $\mathrm{tx}_i$. Waiting tasks become ready as their required state values and preceding updates are resolved.}
  \label{fig:schedule}
  \Description{}
\end{figure*}

\noindent\textbf{Runtime scheduling.} Algorithm~\ref{alg:runtime-scheduling} outlines the runtime task scheduling loop. \sysname{} coordinates symbolic-execution and expression-evaluation tasks by maintaining two ordered task lists \texttt{ReadyTasks} and \texttt{WaitingTasks}. Symbolic-execution tasks of all transactions are initially placed in \texttt{ReadyTasks}, available for concurrent threads to fetch and process. When a symbolic-execution task $E_i$ completes, the scheduler registers the resulting evaluation tasks together with their state dependencies and places them in \texttt{WaitingTasks}. State access or task completions trigger readiness checks for affected waiting tasks with regard to \textit{C1--C3}, allowing eligible tasks to move to \texttt{ReadyTasks}. A symbolic-execution task may also enter the waiting list if it falls back to concrete evaluation during branch prediction. Similar to evaluation tasks, it becomes runnable again once states required to evaluate the branch condition are ready. In \texttt{ReadyTasks}, worker threads fetch tasks in increasing order of transaction index. This prioritization facilitates the progressive resolution of long dependent chains within the block while allowing tasks with satisfied dependencies to execute in parallel.

\noindent\textbf{Example.}
Fig.~\ref{fig:schedule} illustrates expression-granular scheduling across four transactions. In Fig.~5(a), $E_1$ and $E_2$ have completed symbolic execution, producing evaluation tasks $V_1^A$, $V_1^B$, $V_2^B$, and $V_2^C$, which wait for their required states. Meanwhile, $E_3$ is suspended because resolving a state-dependent branch falls back to L4 concrete evaluation and requires the concrete value of $D$, prompting the worker to switch to $E_4$. In Fig.~5(b), the arrival of $B$ enables $V_1^B$ to run, whose update subsequently enables $V_2^B$. These updates proceed even though $V_1^A$ and $V_2^C$ remain blocked on $A$ and $C$, respectively. The arrival of $D$ also allows $E_3$ to complete and generate $V_3^B$ and $V_3^D$. By this point, $E_4$ has also completed, producing $V_4^D$, whose expression depends on both $D$ and $E$. With their inputs ready, $V_3^B$ and $V_3^D$ can be evaluated, advancing the updates to $B$ and $D$ while the remaining state accesses are still pending. In Fig.~5(c), the arrival of $A$, $C$, and $E$ releases the remaining tasks $V_1^A$, $V_2^C$, and $V_4^D$. In particular, $V_4^D$ uses the value of $D$ produced by $V_3^D$ together with the retrieved value of $E$, thereby respecting the dependency between the two transactions. These ready tasks can then execute concurrently, completing the remaining updates in Fig.~5(d).

\noindent\textbf{Misprediction recovery.}
In \sysname{}, branch prediction validation is also performed as an evaluation task for conditional expressions, following the same scheduling process. However, it requires an additional step that checks the validation results against the recorded branch assertions. As \sysname{} schedules tasks optimistically under the assumption that predictions are correct, a failed validation requires recovery of corresponding transactions. Upon detecting a failed validation, \sysname{} invalidates and reschedules the symbolic-execution and expression-evaluation tasks of that transaction. As re-execution may update existing write expressions, introduce new writes, or remove writes that occurred only on the mispredicted path, \sysname{} also invalidates and reschedules tasks of subsequent transactions directly or transitively affected by these changes. To limit recovery costs, \sysname{} prioritizes lower-index ready tasks as this helps detect mispredictions promptly before many subsequent dependent tasks have executed, thus reducing the work required to repair the affected dependency chains.

\subsection{Optimizations}\label{subsec:optimizations}
\noindent\textbf{Merged state access.}
Concurrent workers may request the same state during execution. Although the shared state cache prevents repeated backend reads once a state has been retrieved, it cannot suppress duplicate requests caused by cache misses while the initial read remains in flight. Such duplication becomes more frequent as parallelism and asynchronous reads increase the number of concurrent requests. \sysname{} therefore merges identical in-flight requests through a shared future, which delivers the retrieved value to all requesters once the read completes. Some accesses are indirect, with dynamic targets that depend on prior reads, such as a state key computed by hashing a previously read value. \sysname{} records these target expressions and issues follow-up reads as soon as their required inputs become ready, without waiting for the originating transaction to be scheduled again.

\noindent\textbf{Intermediate result caching.}
Write expressions within a transaction often share intermediate expressions (e.g., $s_B+1$ in Fig.~\ref{fig:symbolic_execution}). Re-evaluating shared expressions from scratch introduces redundant computation. \sysname{} therefore caches materialized intermediate values with compact records of their dependent state versions. When an intermediate expression is revisited, \sysname{} reuses its cached value if the recorded versions match the required input versions. Otherwise, it resumes evaluation from valid cached intermediates and recomputes only the affected suffixes. Through lightweight version checks, this optimization avoids repeated expression-graph traversal, while preventing the reuse of values invalidated by preceding state updates.

\noindent\textbf{Branch repair.}
A branch misprediction normally requires \sysname{} to invalidate and restart the transaction's symbolic-execution and expression-evaluation tasks. However, we observe that some branches are \emph{locally reconvergent}: their paths affect only a bounded sequence of instructions and then rejoin without changing subsequent control flow. For example, the two paths of a Uniswap branch differ only in whether they update two cumulative-price state variables, after which both perform the same reserve update and event emission. \sysname{} identifies such branches using their control flow and symbolic dependencies, and repairs only the affected state updates instead of restarting the entire transaction.

\section{Implementation}\label{sec:implementation}
We implement \sysname{} on Geth v1.17.1 by adding a symbolic interpreter to its execution engine. The interpreter reuses Geth's existing EVM execution framework while minimally extending runtime value containers to preserve symbolic references. Unlike the original interpreter that performs concrete computations over concrete values, the symbolic interpreter propagates unresolved state references and incrementally constructs symbolic expressions.

We implement a per-transaction dynamic table consisting of append-only entries to efficiently maintain symbolic expressions. Reading a state allocates a new entry, and subsequent operations with concrete operands append operation records to its tail. Additional entries are allocated only when operations merge multiple expression chains or attempt to extend an offset already shared by another execution path. This layout forms a space-efficient expression DAG composed primarily of linear operation chains. Stack elements require only an additional lightweight symbolic-reference field $\mathit{entry.offset}$ to locate their corresponding expressions. For byte-oriented containers such as memory, calldata, and return data, \sysname{} maintains lightweight metadata associating symbolic references with the affected byte ranges, while leaving their underlying byte representation unchanged.

For state access, we extend Geth's original \textit{StateDB} to support asynchronous requests, using futures and completion callbacks to deliver retrieved values. We use RocksDB as the key-value store for the state backend and enable its asynchronous I/O path backed by the Linux \texttt{io\_uring} interface to minimize thread blocking during disk access.

\section{Evaluation}\label{sec:evaluation}
Our evaluation aims to answer the following questions:
\begin{itemize}
    \item How well does \sysname{} perform in terms of throughput compared with existing baselines on real-world workloads? (Section~\ref{sec:eval-overall})

    \item How much does each design component contribute to \sysname{}'s overall performance? (Section~\ref{sec:eval-ablation})

    \item How does \sysname{} perform under varying workload characteristics and system configurations? (Section~\ref{sec:eval-factor})

    \item What system resources does \sysname{} utilize to achieve its performance gains?
    (Section~\ref{sec:eval-resource})
\end{itemize}

\subsection{Experimental Setup}\label{subsec:experimentalsetup}
\noindent\textbf{Testbed.} Our experiments are conducted on an Ubuntu 24.04.4 LTS server equipped with a 13th Gen Intel$^\circledR$ Core$\mathrm{^{TM}}$ i7-13700 processor (16 cores, up to 5.2 GHz), 64 GB of memory, and a 1 TB SK hynix NVMe SSD. To ensure fair comparison, all evaluated systems use the same Geth codebase and database configurations. Each reported result averages three independent runs and is verified to produce identical state transitions across all systems.

\noindent\textbf{Baselines.} We compare \sysname{} with four baselines: serial execution, OCC~\cite{icse2022occda}, BlockSTM~\cite{ppopp2023blockstm}, and ParallelEVM~\cite{eurosys2025parallelevm}. The serial baseline represents Geth's native sequential execution. OCC processes transactions in parallel against the initial block state, followed by serial validation and re-execution of conflicting transactions. BlockSTM uses software transactional memory with multi-versioned states to dynamically detect dependencies for re-execution and commit. ParallelEVM follows an approach similar to OCC but leverages static single-assignment (SSA) logs for fine-grained re-execution upon conflicts. We exclude baselines that rely on precomputed static conflict graphs as \sysname{} assumes no prior knowledge of transaction dependencies.

\noindent\textbf{Workloads.}
We evaluate \sysname{} on both real-world and synthetic workloads. The real-world workload replays 50k consecutive Ethereum mainnet blocks (approximately 10M transactions) starting from block height 23,400K (September 2025), using a 102-GiB RocksDB instance containing 298M key-value entries in this block range. The synthetic workload comprises 10M transactions generated following a Zipfian distribution across four representative dApps: ERC-20 token exchange, automated market maker (AMM), collateralized lending protocol, and NFT marketplace. We use the adjustable synthetic workloads to characterize how factors such as workload complexity, block size, access skew, and database settings affect performance of each system.

\begin{figure*}[!t]
  \centering

  \begin{minipage}[t]{0.325\textwidth}
    \vspace{0pt}
    \centering
    \includegraphics[width=\linewidth]{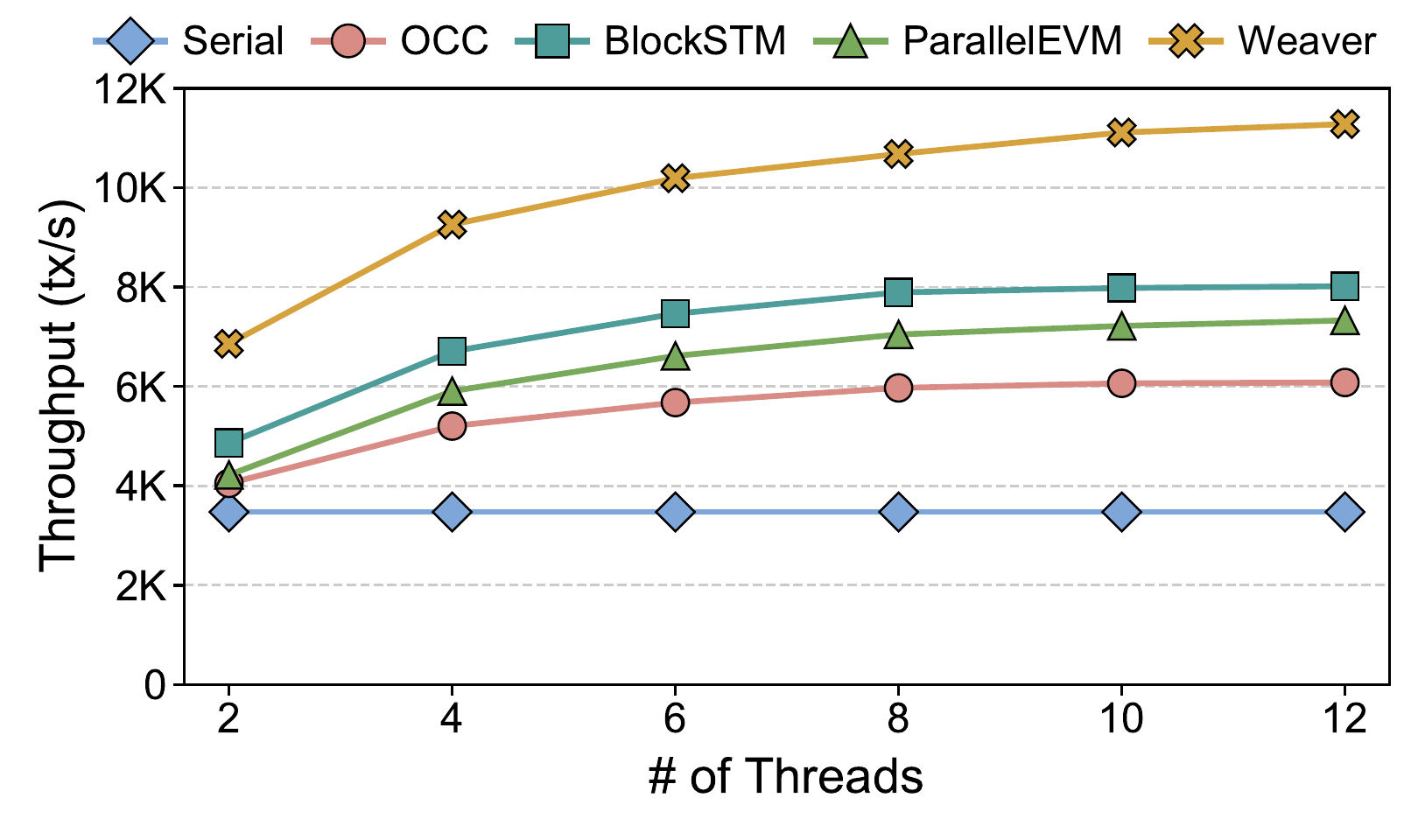}
    \caption{Throughput comparison of different execution engines
      under different thread counts.}
    \label{fig:thread-throughput}
    \Description{}
  \end{minipage}
  \hfill%
  \begin{minipage}[t]{0.65\textwidth}
    \vspace{0pt}
    \centering
    \includegraphics[width=\linewidth]{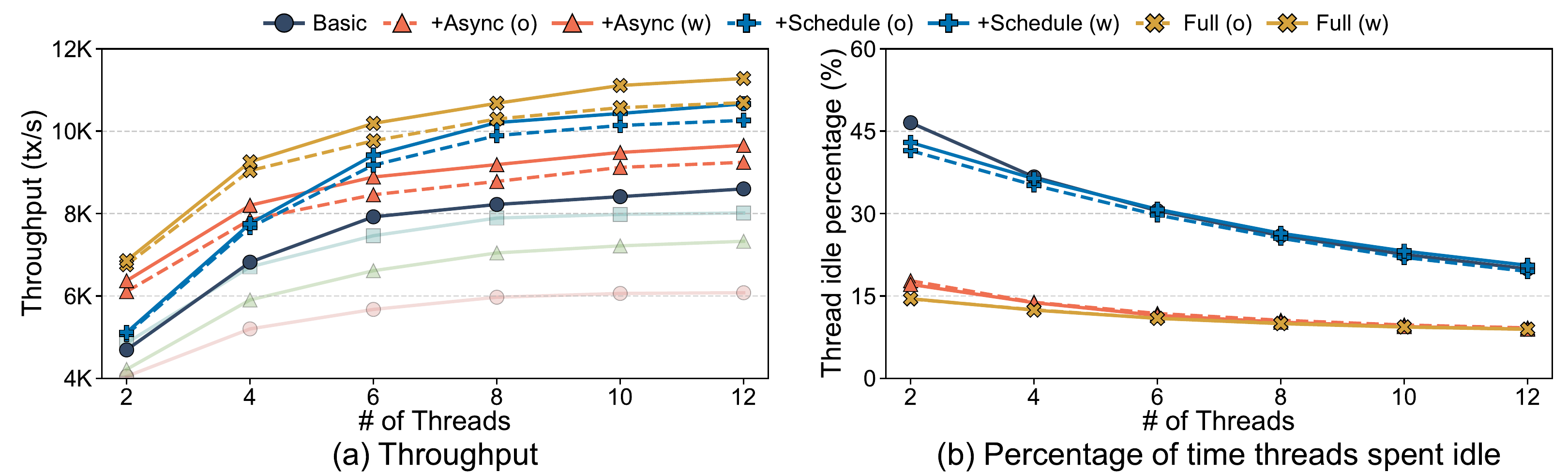}
    \vspace{-5mm}
    \caption{Ablation study of \sysname{} variants. \textit{+Async} and \textit{+Schedule} enable asynchronous state access and expression-granular scheduling, respectively. \textit{Full} combines both mechanisms. Labels w/o denote with/without branch prediction.}
    \label{fig:ablation-study}
    \Description{}
  \end{minipage}

\end{figure*}

\subsection{Overall Performance}\label{sec:eval-overall}
Fig.~\ref{fig:thread-throughput} presents the throughput of \sysname{} and the baselines on the real-world Ethereum workload under different thread counts. While all concurrent execution engines benefit from additional threads, they exhibit different scaling trends. OCC quickly bottlenecks on the serial re-execution of conflicting transactions. ParallelEVM outperforms OCC and delays the saturation by mitigating conflict costs through fine-grained operation redo of conflicting transactions. BlockSTM utilizes multi-versioning that allows transactions to observe and execute based on relatively recent state versions during runtime. It also overlaps validation of preceding transactions with execution of subsequent ones, thus achieving the highest throughput among baselines. Nevertheless, its performance is still capped by the state-blocking execution model. Across all thread counts, \sysname{} consistently delivers the highest throughput. With 12 threads, it reaches 11.28 kTPS, achieving 3.3$\times$ speedup over serial execution and outperforming the best concurrent execution engine by 42\%.

\sysname{}'s performance gains mainly stem from two aspects. First, symbolic execution decouples concrete value access from the critical path of instruction execution, thereby hiding backend latency through asynchronous state access. Second, outputs of symbolic execution, namely the computational expressions associated with each write, naturally decompose a transaction into fine-grained evaluation tasks that can be scheduled independently according to their state readiness. This avoids blocking an entire transaction on isolated pending reads or triggering unnecessary re-executions due to stale states. Overall, \sysname{} fully exploits the parallelism both between I/O and computation and among independent computations within a transaction, whereas the baselines remain bottlenecked by the need for concrete state values and transaction-granular coordination.

\subsection{Ablation Study}\label{sec:eval-ablation}
To isolate the performance impact of each design component, Fig.~\ref{fig:ablation-study} compares four \sysname{} variants. \textit{Basic} employs the core symbolic execution framework, recording state-derived expressions during concurrent execution and selectively re-evaluating affected expressions with updated values during validation. \textit{+Async} adds asynchronous state access, while \textit{+Schedule} adds expression-level scheduling. The \textit{Full} configuration integrates both mechanisms. We also evaluate the latter three variants with (w) and without (o) branch prediction to quantify its impact. Notably, \textit{Basic} alone already outperforms ParallelEVM by 16.2\%. While both systems benefit from partial re-execution, \textit{Basic} incurs lower overhead through its compact state-entry design, which avoids complete SSA construction and facilitates operation tracking across call frames. Furthermore, the branch repair also avoids full transaction re-execution by reconciling localized branch divergences that do not alter the overall control flow.

\noindent\textbf{Effect of asynchronous state access.}
As shown in Fig.~\ref{fig:ablation-study}(a), adding asynchronous state access to \textit{Basic} improves throughput by 12--36\% across all thread counts. While \textit{Basic} waits for concrete values before continuing execution, \textit{+Async} proceeds symbolically while backend reads remain in flight, thereby substantially reducing execution-thread stalls. We further validate this effect by measuring the percentage of time execution threads remain blocked, as detailed in Fig.~\ref{fig:ablation-study}(b). When relying on concrete values to proceed (\textit{Basic} and \textit{+Schedule}), threads spend 20--47\% of their time waiting for state values, whereas decoupled versions (\textit{+Async} and \textit{Full}) reduce this fraction to only 9--17\%. These results demonstrate that asynchronous access effectively removes backend latency from the critical path and allows execution threads to make continuous progress. We also observe that the performance gain is particularly pronounced at low concurrency levels but narrows moderately at higher thread counts. With more transactions executing concurrently, overlapping state requests become more common, allowing the synchronous variant to mask more state-access stalls through shared-cache hits or in-flight request merging.

\noindent\textbf{Effect of fine-grained scheduling.}
On top of \textit{+Async}, \textit{Full} further introduces expression-level scheduling driven by readiness of their respective state dependencies rather than waiting for an entire transaction's read set. As shown in Fig.~\ref{fig:ablation-study}(a), this mechanism improves throughput by an additional 8--18\% across all thread counts. We further find that the average readiness latency of an expression is only 41\% of that of a transaction, with 88\% of expressions becoming ready before their encompassing transactions, indicating unnecessary delays of executable work in transaction-level scheduling. In contrast to the trend observed with asynchronous state access, the \textit{+Schedule} variant provides a smaller benefit at low concurrency but becomes increasingly effective as the thread count grows. This trend emerges because higher concurrency exposes more candidate tasks with diverse readiness states, enabling the scheduler to bypass unavailable computations and keep the execution pipeline populated. Overall, asynchronous state access creates opportunities for overlapping state waits with computation, while fine-grained scheduling converts these opportunities into useful work.

\noindent\textbf{Effect of branch prediction.}
Finally, we evaluate branch prediction on the \textit{+Async}, \textit{+Schedule}, and \textit{Full} variants. As shown in Fig.~\ref{fig:ablation-study}, enabling the predictor improves their throughput by approximately 4.0--5.1\%, 0.9--3.9\%, and 1.5--5.5\%, respectively. For \textit{+Async}, the predictor allows execution to proceed before branch conditions become concrete, avoiding premature dependence on concrete state values. It also benefits \textit{+Schedule}, where expression tasks are generated along the selected branch path. Without prediction, state-dependent branches may follow concrete values later changed by preceding transactions, causing incorrect-path tasks to be scheduled and subsequently recovered. The predictor reduces such incorrect-path scheduling by overriding branch decisions derived from potentially stale concrete values. Tab.~\ref{tab:predictor-comparison} further compares \sysname{} with Seer~\cite{vldb2024seer}. At nearly identical coverage, \sysname{} improves accuracy from 97.0\% to 99.5\%, reduces latency to 16.1\,ns, and nearly halves the model size. Compared with Seer's monolithic perceptron, \sysname{} avoids per-branch history-modeling overhead through its hierarchical architecture. L1 and L2 resolve 57.6\% and 26.1\% of branches with 99.7\% and 99.5\% accuracy, respectively, while L3 covers the remaining complex temporal patterns. This demonstrates that \sysname{} retains high coverage while confining most predictions to lightweight upper levels.

\begin{figure*}[t]
  \centering
  \includegraphics[width=\linewidth]{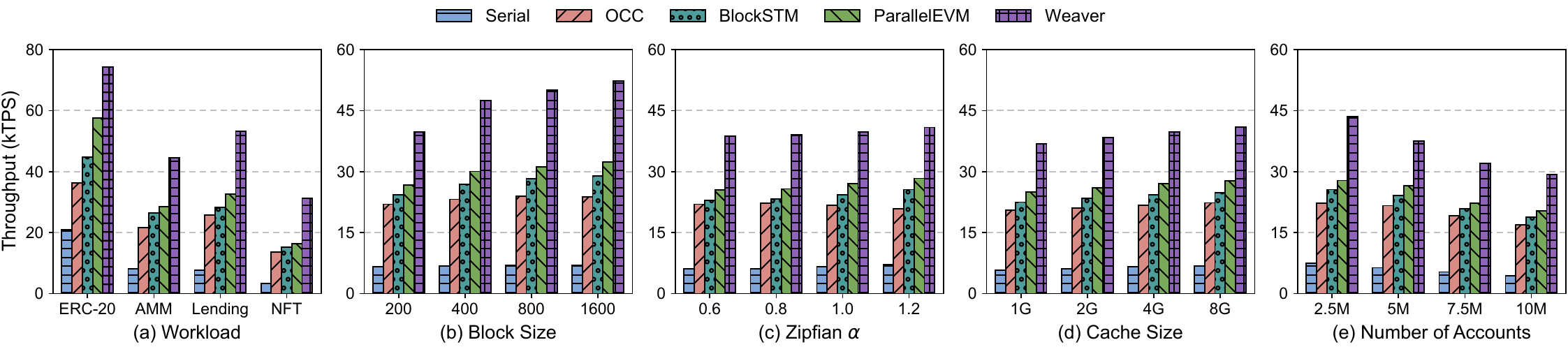}
  \caption{Throughput comparison under different dApps and system configurations. Fig.~(a) evaluates the four dApps individually, while the remaining experiments use a mixed workload comprising all four dApps. Unless varied, the default settings are block size 200, Zipfian $\alpha=1.0$, 4-GiB cache, and 5M accounts.}
  \label{fig:synthetic-throughput}
  \Description{}
\end{figure*}

\begin{table}[t]
  \centering
  \caption{Comparison of branch predictor metrics.}
  \label{tab:predictor-comparison}

  \footnotesize
  \setlength{\tabcolsep}{6pt}
  \begin{tabular}{@{}c|cccc@{}}
    \toprule
    Predictor & Coverage & Accuracy & Latency & Size \\
    \midrule
    Seer~\cite{vldb2024seer}
      & \textbf{88.9\%}
      & \textbf{97.0\%}
      & \textbf{25.2 ns}
      & \textbf{121.6 MiB} \\
    \sysname{}
      & \textbf{88.8\%}
      & \textbf{99.5\%}
      & \textbf{16.1 ns}
      & \textbf{62.4 MiB} \\
    \hspace{2.4em}L1
      & 57.6\% & 99.7\% & 13.7 ns & 8.1 MiB \\
    \hspace{2.4em}L2
      & 26.1\% & 99.5\% & +3.5 ns & 12.3 MiB \\
    \hspace{2.4em}L3
      & 5.1\% & 97.5\% & +5.6 ns & 41.9 MiB \\
    \bottomrule
  \end{tabular}
\end{table}

\subsection{Sensitivity Analysis}\label{sec:eval-factor}
To examine how workload characteristics and system parameters affect transaction execution performance, we conduct a series of sensitivity analyses using synthetic workloads with configurable settings. Fig.~\ref{fig:synthetic-throughput} varies one factor at a time while keeping the others at their default settings.

\noindent\textbf{Impact of workload complexity.}
Fig.~\ref{fig:synthetic-throughput}(a) compares throughput across four application workloads. \sysname{} achieves 74.3, 44.5, 53.3, and 31.2 kTPS on ERC-20, AMM, Lending, and NFT, outperforming the best baseline by 28.9\%, 55.8\%, 63.6\%, and 90.3\%, respectively. ERC-20 features relatively simple logic that allows all systems to achieve high throughput, offering less opportunity for hiding access latency and fine-grained scheduling. In contrast, the latter three dApps entail more intricate logic and broader state interactions, causing synchronous systems to either block execution threads or repair computations based on stale states. \sysname{} overlaps state reads with independent computations and schedules tasks according to state readiness, thereby yielding greater performance benefits as workload complexity increases.

\noindent\textbf{Impact of block size.}
Fig.~\ref{fig:synthetic-throughput}(b) compares system throughput under varying block sizes. As block size increases from 200 to 1600, \sysname{}'s throughput grows from 39.7 to 52.3 kTPS, representing a 31.8\% increase compared with the 20.9\% gain of the best baseline ParallelEVM. Larger blocks increase the density of independent tasks, allowing \sysname{} to better saturate the pipeline and effectively amortize the tail latency of execution or state access observed in smaller blocks. While baselines also benefit from larger blocks, their scalability remains constrained by coarse transaction-level scheduling and synchronous state access on the critical path.

\noindent\textbf{Impact of access skew.}
Fig.~\ref{fig:synthetic-throughput}(c) evaluates different access skews by varying the Zipfian distribution parameter $\alpha$. A larger $\alpha$ concentrates accesses on a smaller set of hot states, increasing both contention and data locality. As $\alpha$ increases from 0.6 to 1.2, OCC throughput decreases as conflict-driven re-executions rise from 0.5\% to 35\%. BlockSTM limits re-execution by operating on more recent state versions, while ParallelEVM reduces the per-transaction re-execution cost. Consequently, their throughput slightly improves as stronger locality offsets the increased contention. \sysname{} maintains the highest throughput across all skew levels, increasing from 38.7 to 40.9 kTPS, as its state-decoupled execution tolerates contention while also benefiting from improved locality.

\noindent\textbf{Impact of database settings.}
Fig.~\ref{fig:synthetic-throughput}(d) and (e) examine the effects of state backend cache size and overall database size, with the latter varied through the number of accounts populated in the database. As the cache size increases from 1 to 8 GiB, all systems benefit from fewer accesses to the underlying database. \sysname{}'s throughput improves from 36.8 to 41.0 kTPS and maintains a stable advantage over the best baseline, demonstrating that its benefits persist beyond cache-constrained environments. Even with higher cache hit rates, asynchronous access continues to hide the residual latency of cache misses, while expression scheduling unlocks greater concurrency that minimizes thread blocking. In contrast, expanding the account space from 2.5M to 10M increases the database size and lowers the cache hit rates, reducing throughput across all systems. \sysname{} maintains 44--54\% higher throughput than the best baseline, demonstrating its robustness across diverse database footprints.

\begin{figure}[t]
  \centering
  \includegraphics[width=\linewidth]{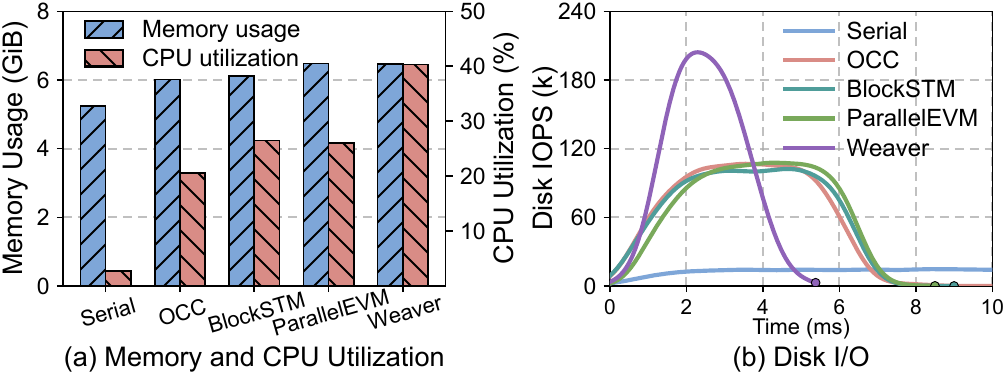}
  \caption{Resource utilization of different transaction execution systems.}
  \label{fig:resource-utilization}
\end{figure}

\subsection{Resource Utilization}\label{sec:eval-resource}
\noindent\textbf{Memory and CPU utilization.}
Fig.~\ref{fig:resource-utilization}(a) compares the memory footprint and CPU utilization of different execution engines. All systems use the same 4-GiB state cache. Serial execution maintains a noticeably smaller runtime working set, while concurrent systems require additional coordination metadata. Among them, OCC introduces the least memory overhead, whereas ParallelEVM and \sysname{} consume moderately more memory. ParallelEVM maintains relatively heavy SSA logs during concurrent execution, while \sysname{} uses compact expression entries and a lightweight branch predictor, keeping its overall memory footprint slightly lower than ParallelEVM. Meanwhile, \sysname{} achieves substantially higher CPU utilization than serial execution and other concurrent baselines, as both asynchronous state access and expression-granular scheduling allow workers to continue useful computation instead of waiting for state values.

\noindent\textbf{Disk I/O utilization.}
Fig.~\ref{fig:resource-utilization}(b) further examines when disk accesses occur during block execution. For each system, we align the disk IOPS curves of 1,000 blocks at block start and average them pointwise, characterizing the typical I/O evolution over a full block-processing cycle (except for Serial). \sysname{} exhibits a notably more front-loaded I/O pattern: its non-blocking execution flow issues outstanding state requests early instead of waiting for preceding accesses to return, producing an early I/O burst and allowing most state retrieval to progress concurrently with transaction execution. This behavior helps \sysname{} complete both state access and block processing earlier than other concurrent engines.

\section{Related Work}\label{sec:relatedwork}
\noindent\textbf{Concurrent transaction execution.} Concurrent transaction execution has been widely studied to exploit multicore resources while preserving serial semantics. Dependency-aware schedulers use conflict information to expose parallel execution opportunities~\cite{icdcs2019parblockchain,icpp2023blockpilot,vldb2026remora}, while optimistic systems relax the need to determine dependencies beforehand by executing speculatively and validating results against the prescribed order~\cite{icse2022occda,sigmod2019fabric,sigmod2020transactional,tkde2021concurrency,vldb2021slimchain}. Subsequent designs reduce conflict costs by narrowing recovery from whole transactions to only the affected computations~\cite{vldb2023hackwrench,eurosys2023morty,vldb2024spectrum,eurosys2025parallelevm}. Multi-version execution instead allows transactions to read more recent state versions during runtime, with validation and re-execution reconciling dependencies as newer preceding versions emerge~\cite{ppopp2023blockstm,icdcs2023dmvcc,osdi2023vmvcc,osdi2024epic}. Recent systems further push parallelism beyond transaction granularity by decomposing computation within contracts~\cite{ppopp2025crystality,middleware2026smartsched}. Despite these advances, existing approaches still commonly rely on synchronous access to concrete state values within transactions or require explicit mechanisms to expose finer-grained execution units. The former leaves state reads as a blocking source along the transaction execution path, while finer-grained execution typically relies on programming-model modifications or declaration of independent computation~\cite{ppopp2025crystality,middleware2026smartsched}. \sysname{} instead decouples execution progress from immediate state availability through symbolic execution and naturally derives expression-level tasks from runtime computational dependencies for fine-grained scheduling and evaluation.

\noindent\textbf{Pre-execution and state-access acceleration.}
Another line of work moves part of transaction processing off the critical path (i.e., ahead to consensus time) and reuses prepared computation or dependency information during formal execution~\cite{sosp2021forerunner,isca2023mtpu,vldb2024seer,nsdi2025vegeta}. State prefetching similarly exploits prior access knowledge to stage required states before replay or execution~\cite{atc2021rainblock,usenixsecurity2026ira,jsa2026mop}. These approaches exploit work completed ahead of execution, and thus depend on an available preparation window, while speculative results may lose value for reuse as the execution context evolves. \sysname{} instead hides state-access latency within the execution path, allowing execution to progress while state requests remain in flight. At another layer, prior work reduces state-access overhead through more efficient authenticated data structures~\cite{osdi2023lvmt,eurosys2026ethane,sigmod2026swmt} or blockchain-tailored storage backends~\cite{eurosys2026ficusdb,sigmod2024chainkv,tcad2024blocklsm}. Both classes of techniques are largely orthogonal to \sysname{} and can complement its runtime decoupling to further improve overall system performance.

\noindent\textbf{Symbolic execution for smart contracts.}
Symbolic execution has been widely used in smart-contract security analysis to explore feasible executions and expose vulnerable program behaviors~\cite{ccs2016oyente,acsac2018maian,usenixsecurity2018teether,usenixsecurity2020ethbmc,usenixsecurity2021smartest,usenixsecurity2023confusum}. Beyond vulnerability analysis, it has also supported formal reasoning about contract semantics~\cite{cav2023evmoptimization,usenixsecurity2023fasverif} and guided testing toward behaviors of interest~\cite{ccs2019ilf,eurosp2021confuzzius,sp2022sailfish}. Overall, these efforts primarily use symbolic execution as an offline program-analysis technique, where relatively detailed recording and extensive path exploration can be tolerated. \sysname{} instead embeds symbolic execution into runtime transaction processing, with lightweight symbolic representations and single-path prediction to make it suitable for latency-sensitive execution.

\section{Conclusion}\label{sec:conclusion}
This paper presents \sysname{}, a state-decoupled blockchain execution engine that uses lightweight symbolic methods to decouple transaction execution from concrete state retrieval. Its hierarchical branch prediction handles state-dependent control flow, while expression-granular scheduling advances individual computations as their state inputs become ready, exposing finer-grained parallelism and reducing execution stalls. Evaluation shows that \sysname{} substantially improves throughput over state-of-the-art execution engines.

\bibliographystyle{ACM-Reference-Format}
\bibliography{reference}

\end{document}